%% file: jfgo.tex
\documentclass{PoS}

\usepackage{units}
\usepackage[utf8]{inputenc}
\usepackage{wrapfig}

\title{Emergence of Quark--Gluon Plasma Phenomena}

\ShortTitle{Emergence of Quark--Gluon Plasma Phenomena}

\author{\speaker{Jan Fiete Grosse-Oetringhaus}\\
CERN, 1211 Geneva 23, Switzerland\\
E-mail: \email{jgrosseo@cern.ch}}

\abstract{The discovery of QGP phenomena in small collision systems like pp and p--Pb collisions have challenged the basic paradigms of heavy-ion and high-energy physics. These proceedings give a brief overview of the key findings and their implications, as well as today's experimental and theoretical situation. An outlook of future measurement is made.}

\FullConference{
European Physical Society Conference on High Energy Physics - EPS-HEP2019 -\\
10-17 July, 2019\\
Ghent, Belgium}

\include{commands}

\begin{document}

\section{Introduction}

Heavy-ion collisions at ultrarelativistic energies allow to produce a Quark--Gluon Plasma (QGP) in the laboratory. The QGP is expected to have prevailed in the early universe and its study allows to access the regime of deconfined quarks and gluons.
With the measurements at the LHC at CERN and at RHIC at BNL, the last decade has advanced the field into the precision era, both, for the wealth of observables and the unprecedented quantification of the observed effects~\cite{Schukraft:2017nbn}. In addition to the detailed study of Pb--Pb and Au--Au collisions, smaller systems (p--Pb and d--Au collisions) as well as pp collisions are investigated, initially as a reference for the measurements in large systems.
Surprising discoveries have been made in these smaller systems which have shaken the basic paradigm of the field of heavy ions. This basic paradigm assumes that the phenomena observed in heavy-ion collisions requires the formation of a QGP. In turn the formation of a QGP requires a large enough volume of hot and dense matter and therefore collisions of large objects. The experimental evidence discussed in this write-up questions this paradigm.
The discoveries have led to a tremendous experimental and theoretical activity in recent years: some of the related publications rank among the highest cited publications of the ALICE, ATLAS and CMS collaborations~\cite{Abelev:2012ola,Aad:2012gla,Khachatryan:2010gv}. A selection of results will be reviewed in the following, but it can be hardly given justice to the overall activity here due the space limitation. For a full review, the reader is invited to consider Ref.~\cite[Chapter 9]{Citron:2018lsq} and Ref.~\cite{Loizides:2016tew}.

\begin{figure}[b]
\centering
\includegraphics[width=0.49\linewidth,trim=300 20 0 290,clip=true]{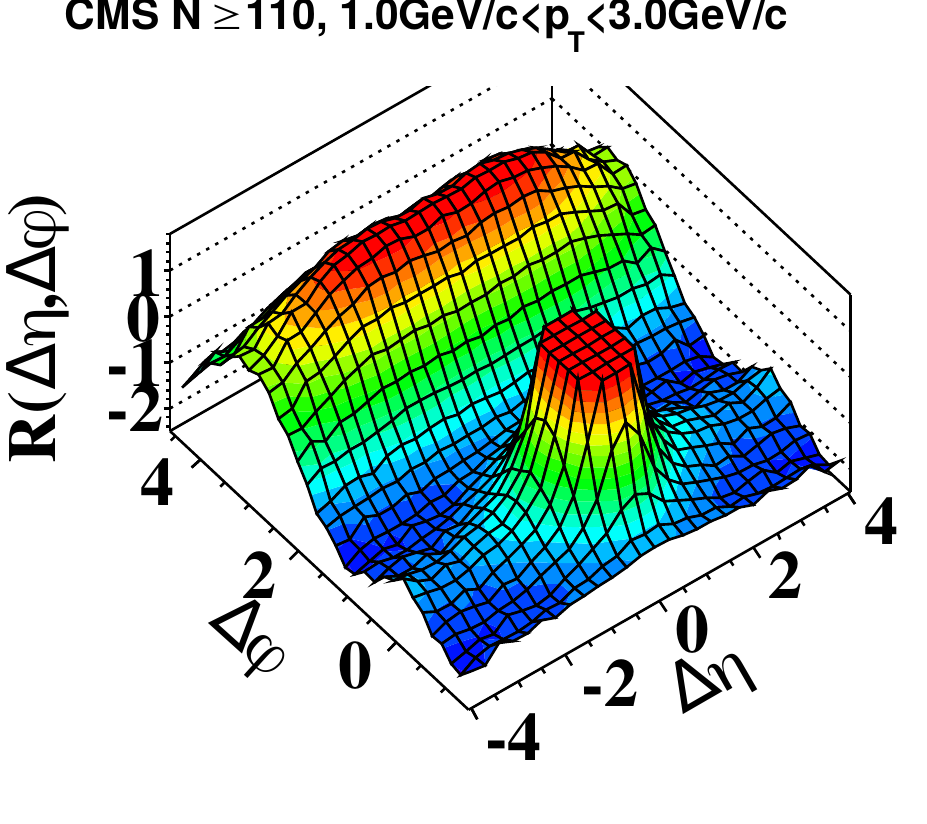}
\hfill
\includegraphics[width=0.49\linewidth]{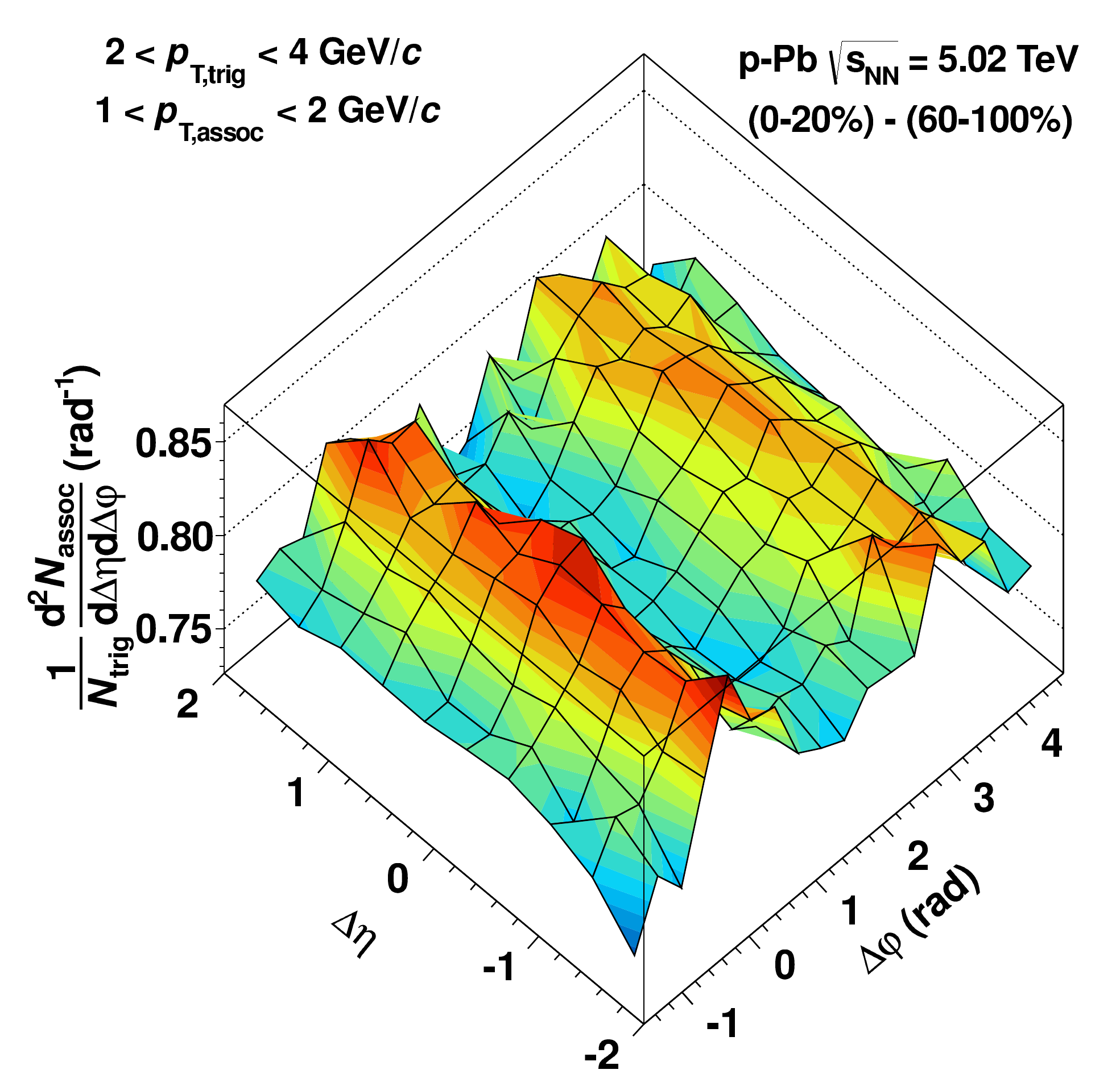}
\caption{Two-particle correlations in high-multiplicity pp collisions (left panel, figure from Ref.~\cite{Khachatryan:2010gv}) and high-multiplicity p--Pb collisions after subtraction of the low-multiplicity counterpart (right panel, figure from Ref.~\cite{Abelev:2012ola}).}
\label{fig:ridge}
\end{figure}

In retrospect, the discovery of long-range correlations in two-particle correlations in very high multiplicity pp collisions~\cite{Khachatryan:2010gv} marked the beginning of the study of emerging QGP phenomena. Correlations of particles with a transverse momentum of a few \unit{GeV/$c$} are dominated by a so-called near-side peak structure and an away-side ridge structure. Both originate in the fragmentation of a $2 \rightarrow 2$ parton scattering process into hadrons. 

While the particles on the near side are found within small angular difference both in azimuth and in pseudorapidity (1--2 units depending on the momenta of the involved particles), the away side retains only the correlation back-to-back in azimuth, due to the fact that the center-of-mass system of the scattering is not corresponding to the lab frame in hadronic collisions. In Ref.~\cite{Khachatryan:2010gv}, an additional correlation is observed at large pseudorapidity differences on the near side for very high multiplicity collisions which was named \emph{the ridge}, see Fig.~\ref{fig:ridge} (left).
\begin{wrapfigure}{Rb!}{0.5\textwidth}
\includegraphics[width=\linewidth]{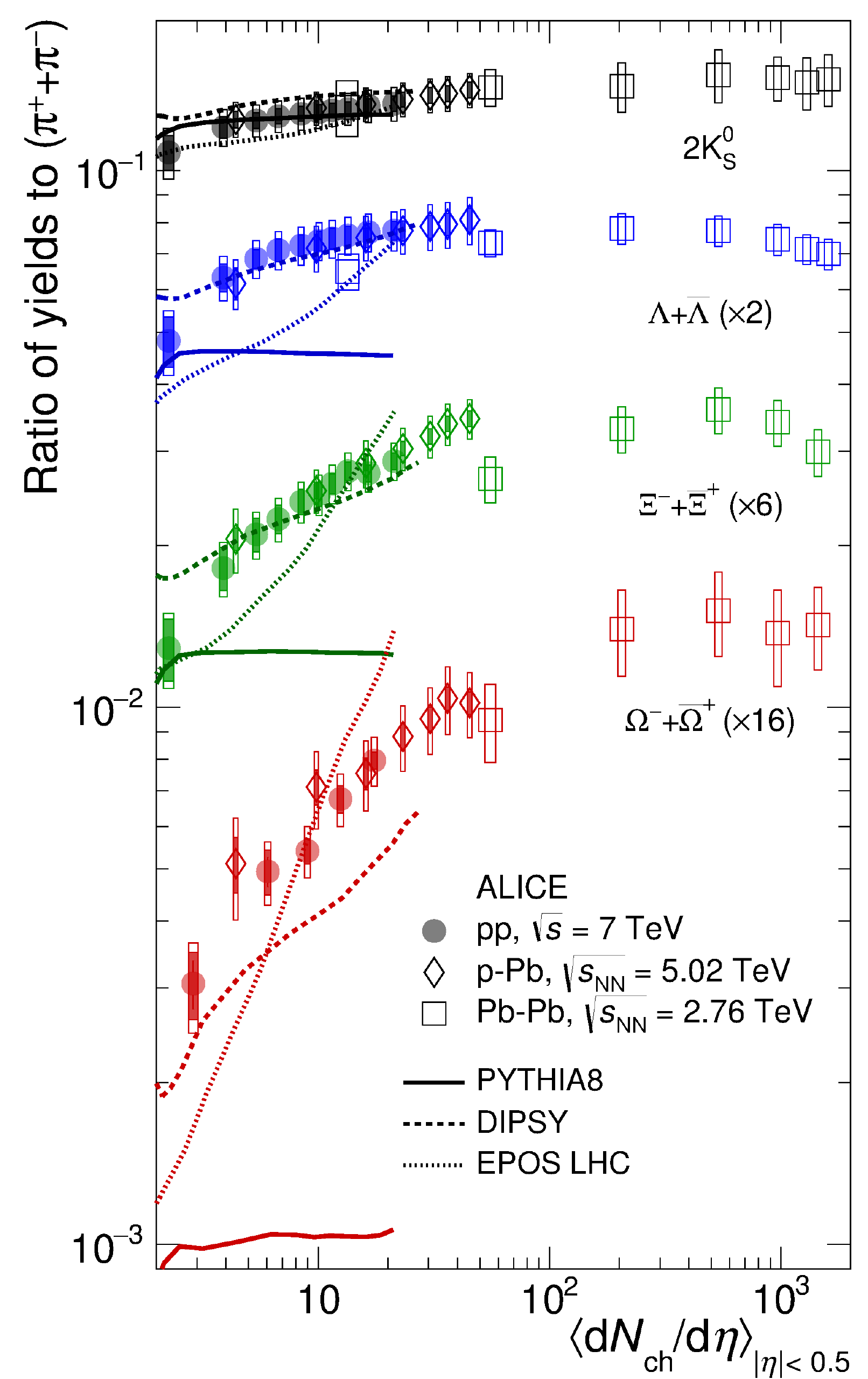}
\caption{Ratios of the strange particles $K^0_S$, $\Lambda$, $\Xi$, and $\Omega$ to pions as a function of charged-particle multiplicity. Results from pp, p--Pb and Pb--Pb collisions are shown, overlaid by model comparisons. 
Figure from Ref.~\cite{ALICE:2017jyt}.}
\label{fig:strangeness}
\end{wrapfigure}
A similar structure was observed in p--Pb collisions~\cite{CMS:2012qk} accompanied by a further ridge structure on the away side, see Fig.~\ref{fig:ridge} (right), made visible employing a subtraction procedure~\cite{Abelev:2012ola,Aad:2012gla}. While the first ridge structure already reminded of so-called elliptic flow observed in heavy-ion collisions~\cite{Aamodt:2010pa}, the observation of two ridge structures made this reminiscence indisputable. In heavy-ion collisions this phenomenon is directly attributed to the hydrodynamic expansion of the hot and dense matter~\cite{Ollitrault:1992bk}.

The second surprising observation is that strange baryon production increases faster than multiplicity~\cite{ALICE:2017jyt}. Increased strangeness production has been observed in large systems and is seen traditionally as a sign of deconfinement as it is energetically cheaper to produce a pair of strange quarks than a pair of strange hadrons. Surprisingly, the increased strangeness production is already present in pp collisions when studying higher multiplicity and connects smoothly to p--Pb and then Pb--Pb collisions. Figure~\ref{fig:strangeness} presents particle ratios for four strange particle species as function of multiplicity. One observes that traditional MC codes, like PYTHIA~\cite{Sjostrand:2007gs}, completely fail to reproduce the trend which has been identified as a significant conceptual problem in such models~\cite{Sjostrand:2018xcd}. These discoveries triggered a large experimental programme as well as significant theoretical modelling which is briefly reviewed in the following section.

\section{Experimental Situation Today}

The ridge structures shown in Fig.~\ref{fig:ridge} are quantified in detail by their Fourier coefficients $v_n$ of the azimuthal distribution, defined as:
\begin{equation}
\frac{dN}{d\varphi} \propto 1 + 2 \sum_n v_n \cos n(\varphi - \Psi_n),
\end{equation}
where $\varphi$ is the azimuthal angle of the particle and $\Psi_n$ the n$^{\rm th}$s order participant plane~\cite{Alver:2010gr}.
In A--A collisions, the dominant component is the second-order coefficient $v_2$ called \emph{elliptic flow}, primarily driven by the elliptic shape of the overlap between the two colliding nuclei.
However, also higher-order ($n>2$) components have a significant contribution which showed that the internal structure of the initial matter distribution in the colliding particles of the nuclei needs to be considered~\cite{ALICE:2011ab}. The fluctuating positions of the nucleons in the nuclei lead to a different matter distribution event-by-event. 
These anisotropies of the initial matter distribution lead to asymmetries in the final-state momenta, when sufficient interactions between the constituents occur. This transition can be described by hydrodynamic models which treat the QGP as a liquid with certain properties which can then be extracted by comparison of the measured $v_n$ coefficients to theoretical calculations. 
Given that the underlying symmetry planes are determined by the initial state of the collisions, they are identical for all outgoing particles and therefore all particles are correlated with each other. This has to be separated from few-particle correlations stemming from jets or resonance decays.

\begin{figure}[t]
\centering
\includegraphics[width=\linewidth]{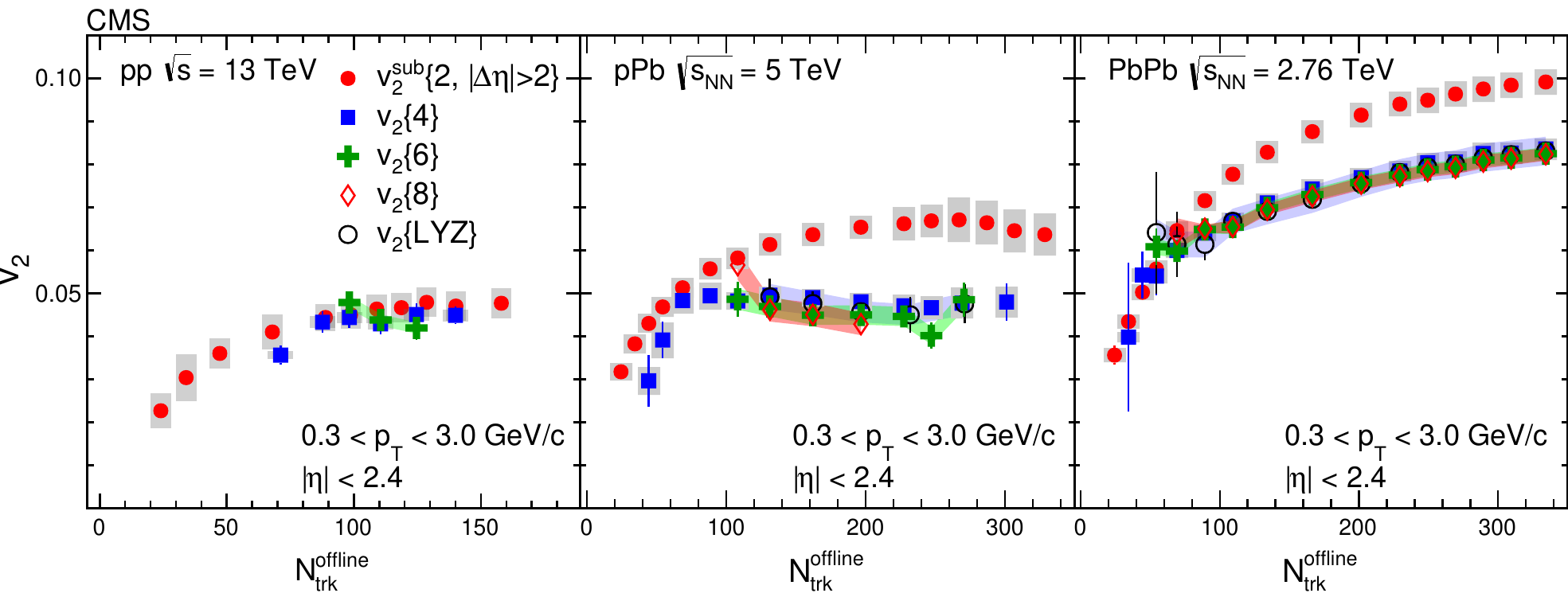}
\caption{The coefficient $v_2$ as a function of multiplicity in pp (left), p--Pb (center) and Pb--Pb collisions (right). A significant value is seen for all collision systems for measurements with up to 8 particles. Figure from Ref.~\cite{Khachatryan:2016txc}.}
\label{fig:v2}
\end{figure}

Given the much smaller overlap region in pp and p--Pb collisions, the measurements of significant components in these collisions came as a surprise. Figure~\ref{fig:v2} shows $v_2$ measurements in pp, p--Pb and Pb--Pb collisions as a function of multiplicity. In order to exclude that jet-like correlations have significant influence on the measured coefficients, multi-particle correlation techniques are used showing that the observed effects involve at least 6 (8) particles in pp (p--Pb and Pb--Pb) collisions~\cite{Khachatryan:2016txc}. The overall magnitude is similar in pp and p--Pb collisions and somewhat smaller than in Pb--Pb collisions. Generally, the similarities are striking.

\begin{figure}[t]
\centering
\includegraphics[width=0.46\linewidth]{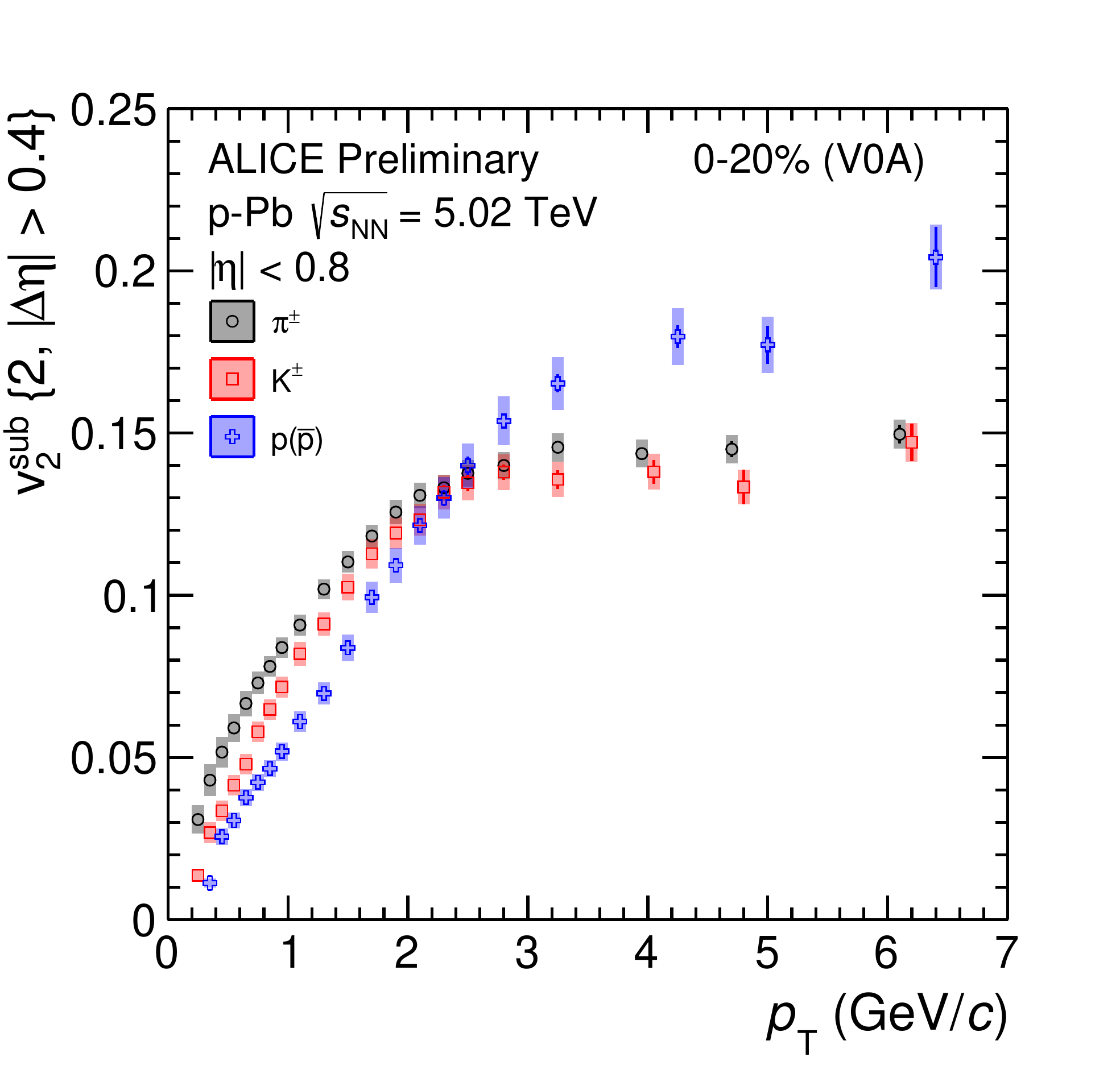}
\hfill
\includegraphics[width=0.52\linewidth]{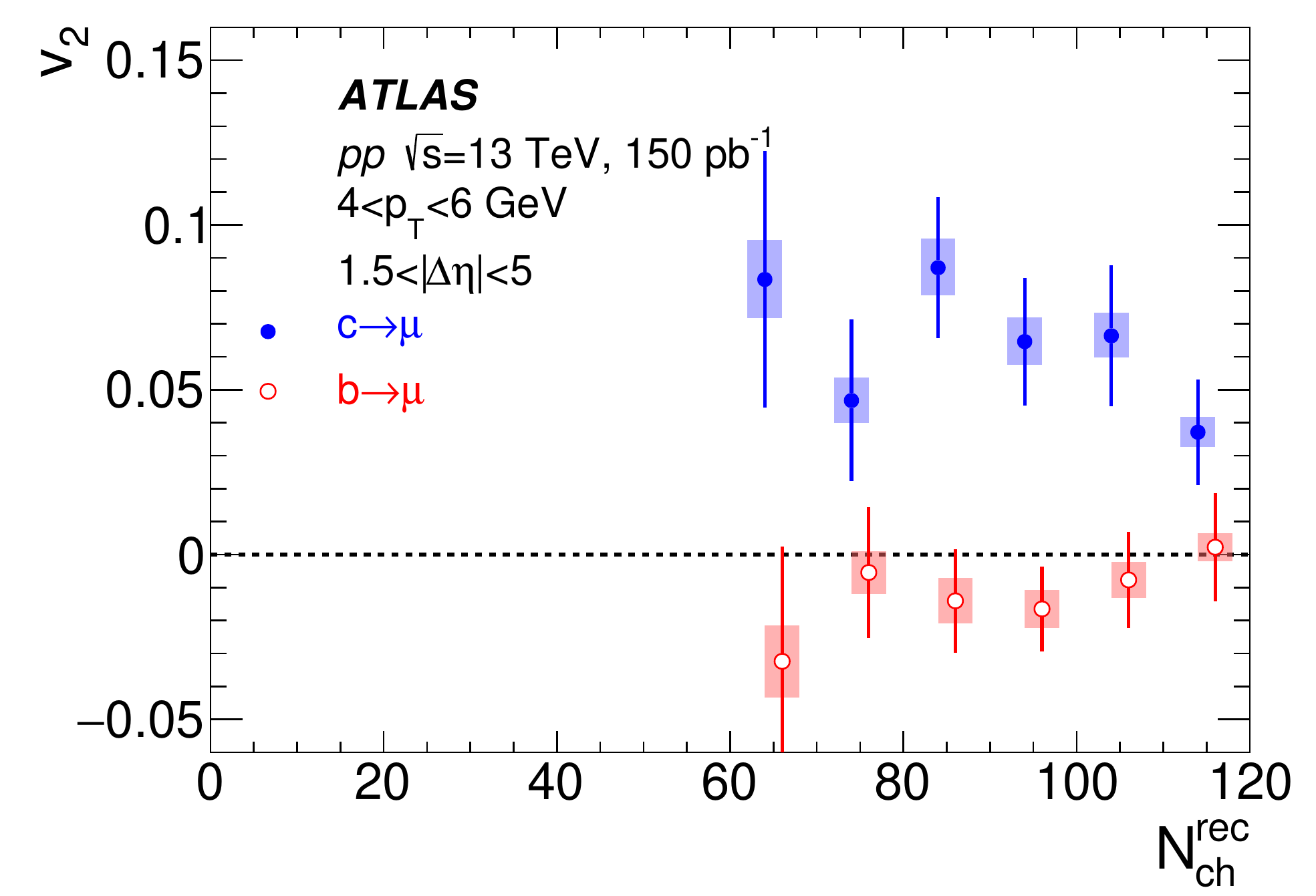}
\caption{Left: $v_2$ coefficient for $\pi$, K, and p as a function of $\pt$ in p--Pb collisions. A characteristic splitting and crossing of the $v_2$ of the different particle species is observed. Right: $v_2$ coefficient as a function of multiplicity for $b$ and $c$ heavy-flavour decay muons. A positive value is observed for charm quarks, while the result for $b$ is consistent with 0. It should be noted that the measurement is for $4 < \pt < \unit[6]{GeV/\emph{c}}$ and thus does not include the low-momentum region. Figure from Ref.~\cite{Aad:2019aol}.}
\label{fig:pid}
\end{figure}

The large energy density of the hot and dense matter gives rise to a common velocity field with which the constituents of the medium rapidly expand. A consequence of this so-called \emph{radial flow} is that all particles have a similar $\beta$ resulting in a mass-dependent influence on the particle momenta. This well-known effect from Pb--Pb collisions has also been observed in pp and p--Pb collisions, see the left panel of Fig.~\ref{fig:pid}, providing further evidence for an expanding medium also in these small collision systems. Further insight can be obtained from the study of heavier charm and beauty quarks. Figure~\ref{fig:pid} (right) shows the measurement of heavy-flavour decay muons from charm and beauty decays~\cite{Aad:2019aol}. This result and additional measurements involving heavy-flavour decay electrons~\cite{Acharya:2018dxy} as well as $D$ and $J/\Psi$ mesons~\cite{Sirunyan:2018kiz} show that also the charm quark has a significant $v_2$ component. For the $b$ to date, no signal has been seen at large $\pt$. While this could indicate that the beauty quark is too heavy to participate in the system evolution, the low-momentum region remains to be studied before a final answer can be given.

\begin{figure}[t]
\centering
\includegraphics[width=0.47\linewidth]{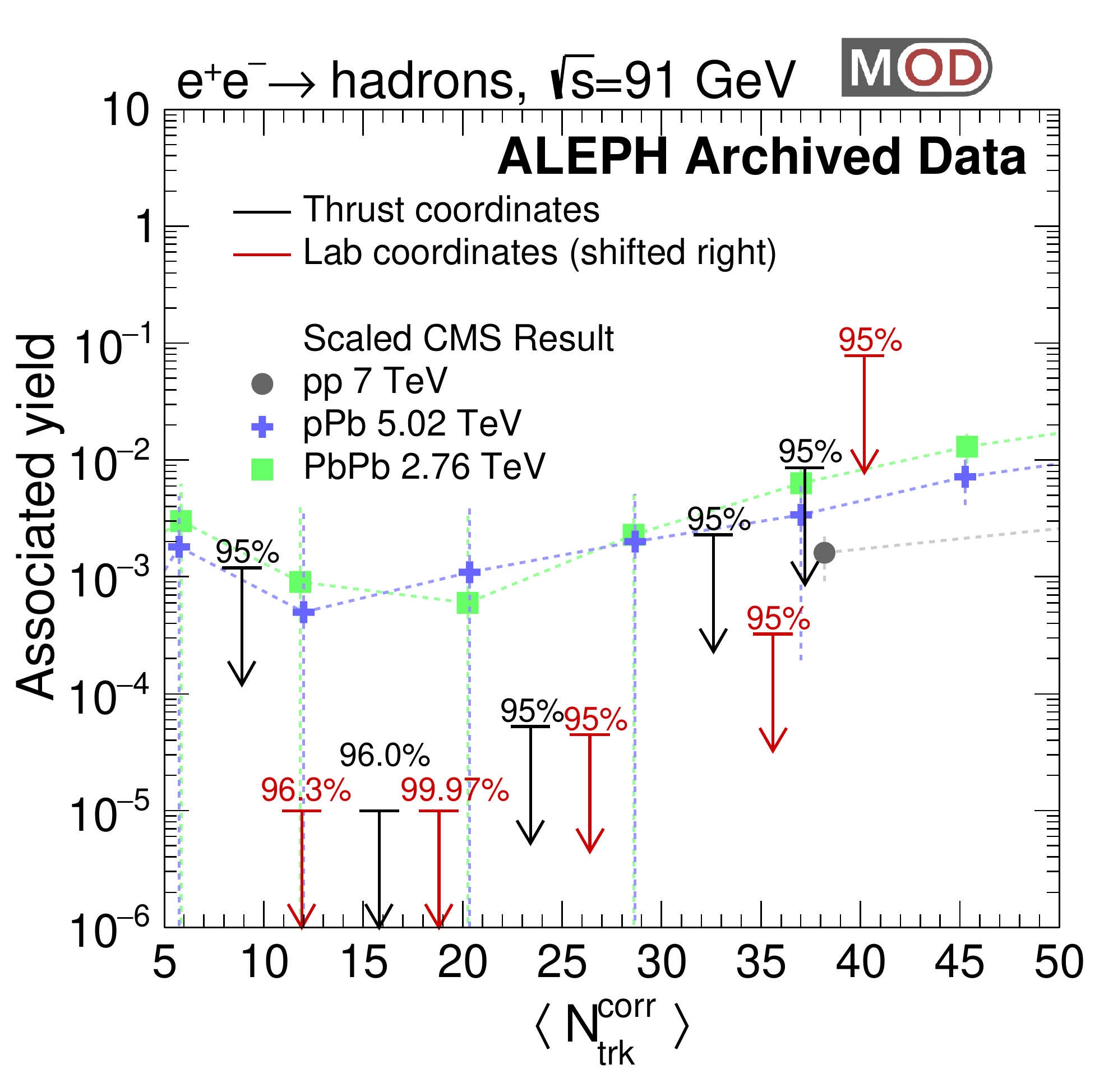}
\includegraphics[width=0.45\linewidth,clip=true,trim=0 0 340 0]{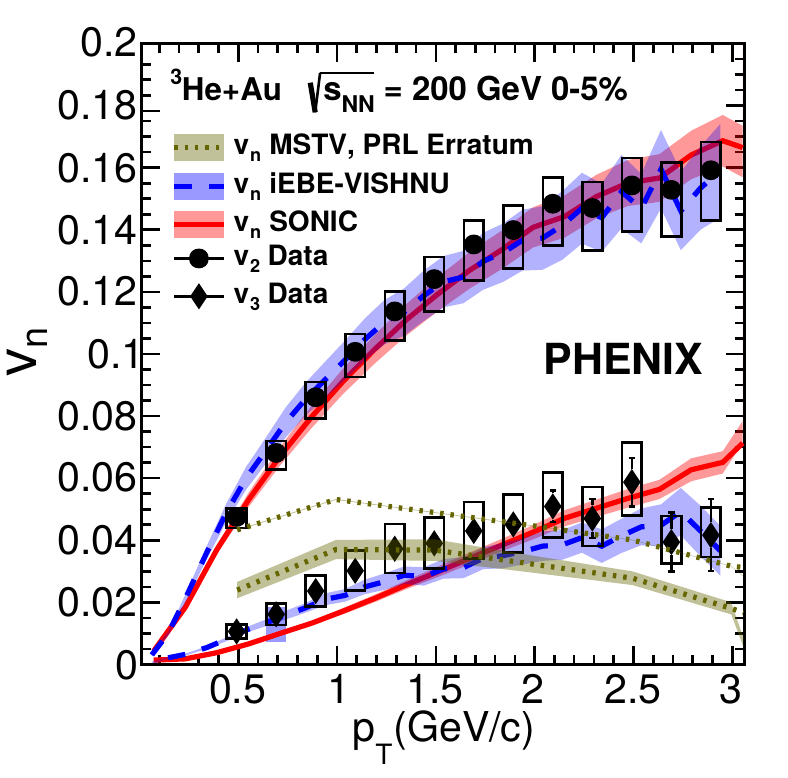}
\caption{Left: limits on near-side ridge yields in $e^+e^-$ collisions recorded by ALEPH compared to results from the LHC. For a detailed discussion of the comparison see text. Figure from Ref.~\cite{Badea:2019vey}. Right: $v_2$ and $v_3$ as a function of $\pt$ in $^3$He--Au collisions compared to hydrodynamic calculations (iEBE-VISHNU and SONIC) which predicted the values well. The calculation involving only initial-state momentum correlations (MSTV) cannot reproduce the measurement. Figure adapted from Ref.~\cite{PHENIX:2018lia}.}
\label{fig:aleph_phenix}
\end{figure}

In order to investigate if the observed ridge structures could be related to a fundamental process and thus not need any final-state interactions, archived $e^+e^-$ collisions recorded by ALEPH have been analyzed. No signal has been observed and Fig.~\ref{fig:aleph_phenix} (left) compares the obtained limit with results from pp, p--Pb and Pb--Pb collisions. At multiplicities below 30, the limit on the associated yield is about $10^{-5}$, while the uncertainties in hadronic collision systems are of the order of $10^{-3}$. At larger multiplicities, the signal observed in hadronic systems is finite but compatible with the (poorer) limit in $e^+e^-$ collisions. While multiplicities between the systems may not be directly comparable, the call is still out if there is a significant difference between $e^+e^-$ and hadronic collisions.

The versality of RHIC allowed a detailed comparison of collisions with different shapes of the initial overlap region. This has been achieved by colliding p, d, and $^3$He on Au. While p--Au is rather round, d--Au and $^3$He--Au have a large elliptic component. In addition, $^3$He--Au has on average about twice the triangularity than the other two systems. The performed measurements in these collisions, show that the shape of the initial state determines the strength of the measured $v_2$ and $v_3$ coefficients~\cite{PHENIX:2018lia}. This has important consequences: the transition from the initial-state shape to the final-state momenta requires interactions of the constituents. Furthermore, hydrodynamic models implementing those correctly predicted the measured values~\cite{Habich:2014jna,Shen:2016zpp} while models involving only initial-state momentum correlations cannot reproduce the effects~\cite{Mace:2018vwq}. Figure~\ref{fig:aleph_phenix} (right) presents these coefficients in $^3$He--Au collisions compared to these model calculations.

\begin{figure}[t]
\centering
\includegraphics[width=0.52\linewidth]{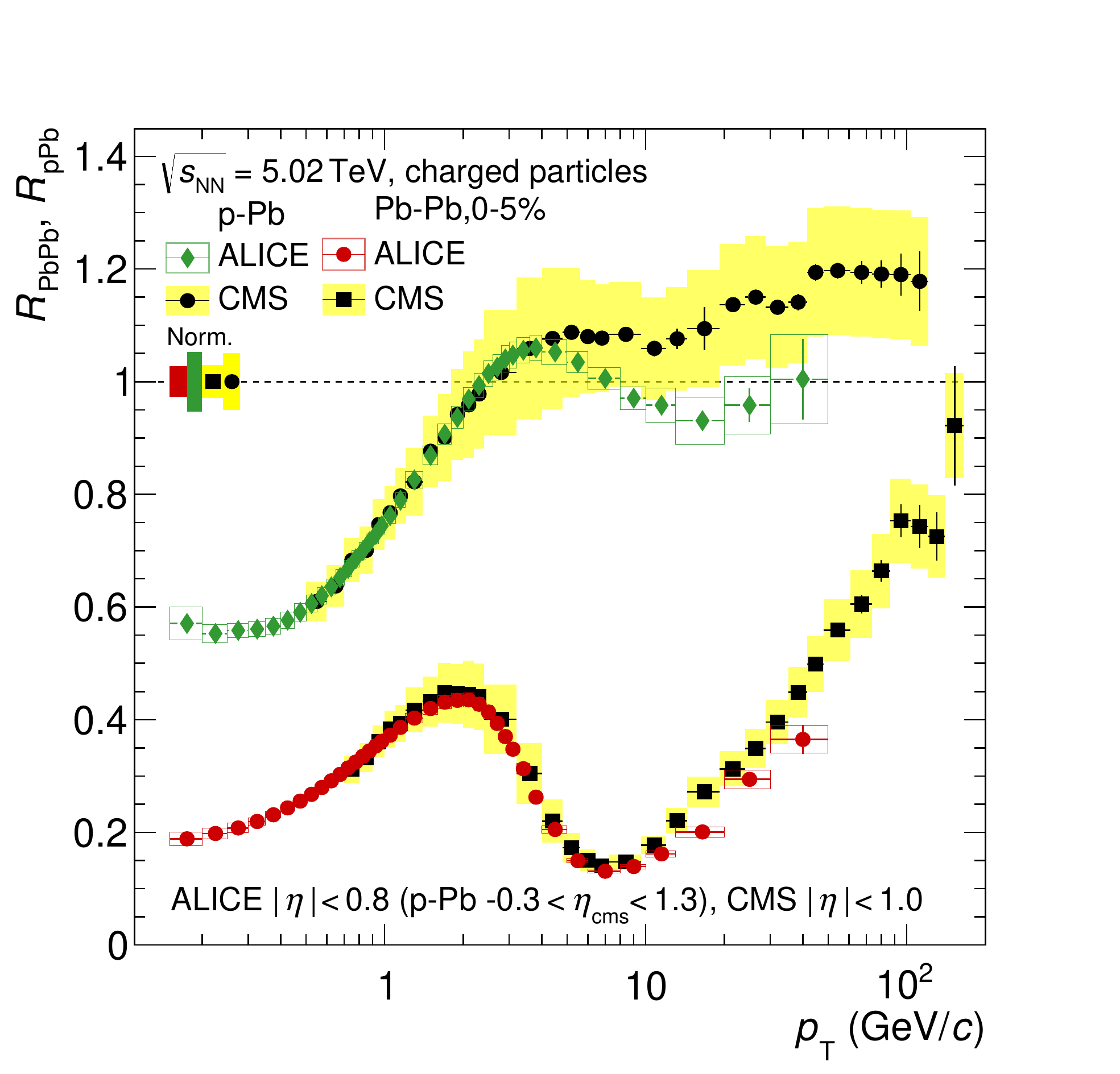}
\hfill
\includegraphics[width=0.46\linewidth]{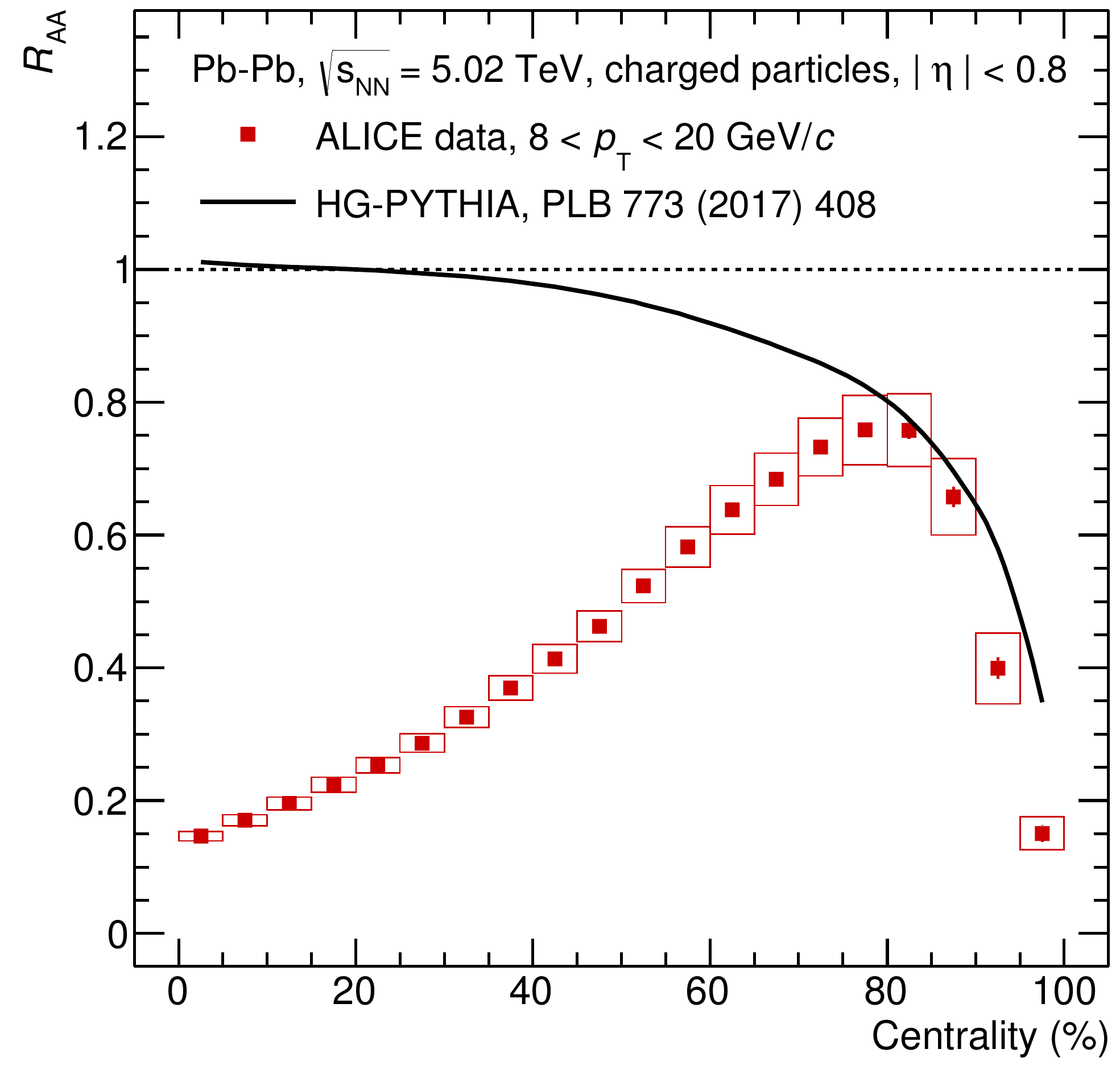}
\caption{Left: $\RAA$ and $\RpA$ as a function of $\pt$ from ALICE and CMS. Figure from Ref.~\cite{Acharya:2018qsh}. Right: $\RAA$ for $8 < \pt < \unit[20]{GeV/c}$ as a function of centrality compared to a simple superposition model not involving parton energy loss. Figure from Ref.~\cite{Acharya:2018njl}.}
\label{fig:energyloss}
\end{figure}

As discussed, the presented results support the idea of final-state interactions in small collision systems. If such interactions are indeed present, the outgoing partons should also lose energy by this mechanism. This phenomenon is well known from collisions of large systems where high $\pt$ hadrons and jets lose a significant fraction of their energy~\cite{Acharya:2018qsh}. However, in small systems a signal of parton energy loss has not been observed to date, neither for inclusive hadrons~\cite{Acharya:2018qsh,Khachatryan:2015xaa} (see Fig.~\ref{fig:energyloss}, left panel), nor jets~\cite{Adam:2016jfp,Sirunyan:2016fcs}, nor D mesons~\cite{Aaij:2017gcy,Acharya:2019mno}, nor B and $J/\Psi$ from B~\cite{Aaboud:2018quy,Aaij:2019lkm}. Also h--jet coincidence measurements can only provide an upper limit on parton energy loss in p--Pb collisions~\cite{Acharya:2017okq}. This creates an apparent inconsistency as the well-established observable $\RAA$ showed a difference from unity (the expectation if A--A collisions were an incoherent superposition of nucleon--nucleon collisions) in peripheral collisions. The latter are at similar multiplicities where unity was measured in p--Pb collisions. This inconsistency was recently understood by a measurement in very peripheral Pb--Pb collisions (80--100\%) where an unphysical reduction of $\RAA$ was observed~\cite{Acharya:2018njl} and explained by a simple superposition model~\cite{Morsch:2017brb}, see Fig.~\ref{fig:energyloss} (right). This model includes the variation of the impact parameter of the single nucleon--nucleon collisions and its effect on the event classification. In consequence, signals of parton energy loss seem to be absent in peripheral Pb--Pb collisions and p--Pb collisions, although the presence of final-state interactions should give rise to them to some extent.

In addition, to this puzzling absence of energy loss in small systems, on open question is the magnitude of $v_2$ coefficients at low multiplicity in pp collisions. Their extraction in low-multiplicity collisions is very challenging due to dominating jet-like correlations and resonance decays. Depending on the utilized subtraction method a finite~\cite{Aaboud:2017acw} or close to zero~\cite{Khachatryan:2016txc} $v_2$ is extracted in low-multiplicity pp collisions. The fact that the result is procedure-dependent, means that the collective nature in dilute systems is not understood, yet.

\section{Explanations \& Modelling}

\begin{figure}[t]
\centering
\includegraphics[width=\linewidth]{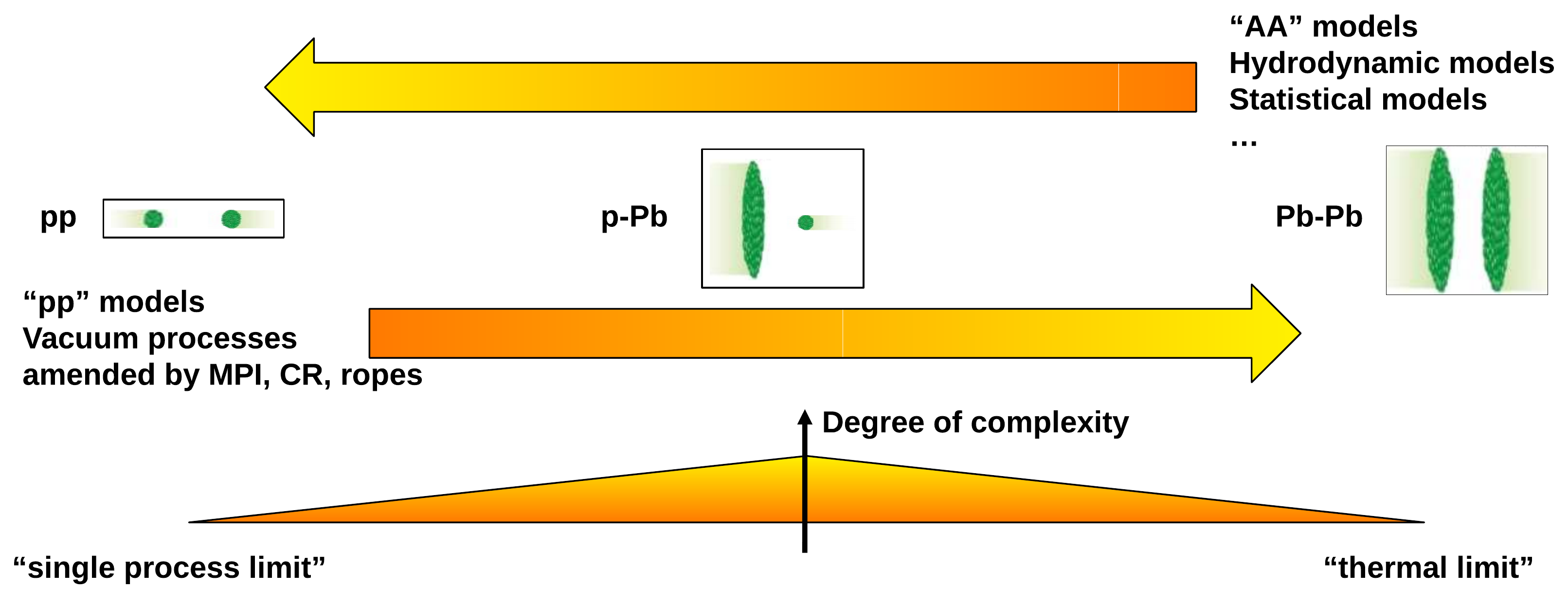}
\caption{Illustration of the landscape of modelling from pp collisions in the single-process limit to Pb--Pb collisions in the thermal limit. Models are extended along both directions, where the degree of complexity increases going away from one of the limits.}
\label{fig:modelling}
\end{figure}

The observations of QGP phenomena in small systems have received wide attention. Their theoretical explanation and description attempts can be grouped into three areas:
\begin{itemize}
  \item Extending the hydrodynamic description valid in large collision systems involving many constituents to small systems. This approach assumes many scatterings between the constituents.
  \item An approach showing that few scatterings can already create anisotropies called \emph{escape mechanism}.
  \item Considering the effect of momentum correlations in the initial state of the colliding objects. In this approach, no final-state interactions are considered, although it can be combined with the other approaches.
\end{itemize}
These three areas span the entire field between fluid dynamics (many scatterings) and the free-streaming limit (no scatterings).
Figure~\ref{fig:modelling} illustrates the two modelling directions which follow. The first approach starts from a valid description in large systems (for example hydrodynamics or statistical models), and extends it in the direction of smaller systems. The second approach begins with a valid description in vacuum (for $e^+e^-$) possibly amended by multiple parton interactions, color reconnection and ropes (for pp) and extends it to larger systems. In both approaches the degree of complexity increases when moving towards intermediate systems like p--A collisions.

\begin{figure}[t]
\centering
\includegraphics[width=0.54\linewidth]{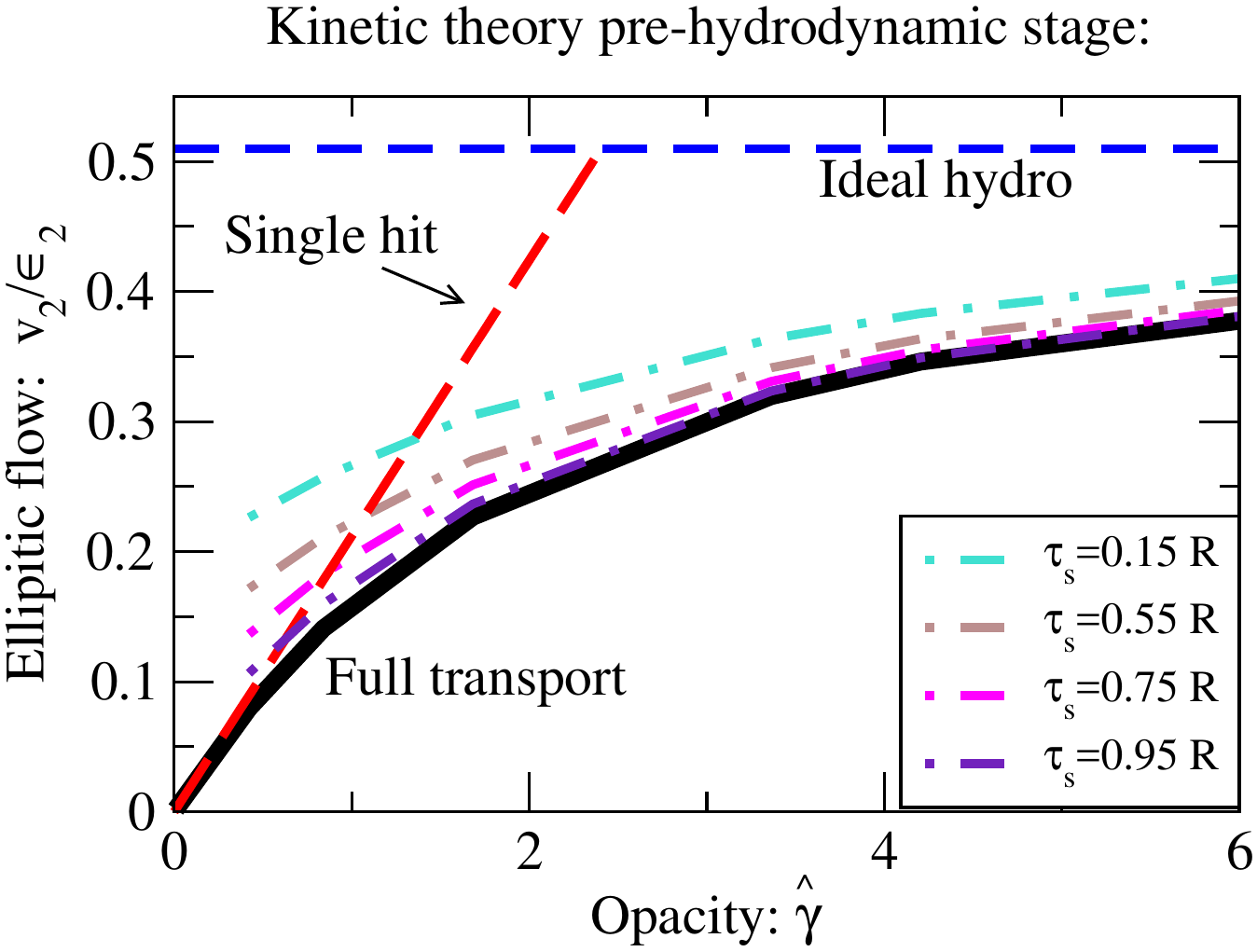}
\hfill
\includegraphics[width=0.44\linewidth]{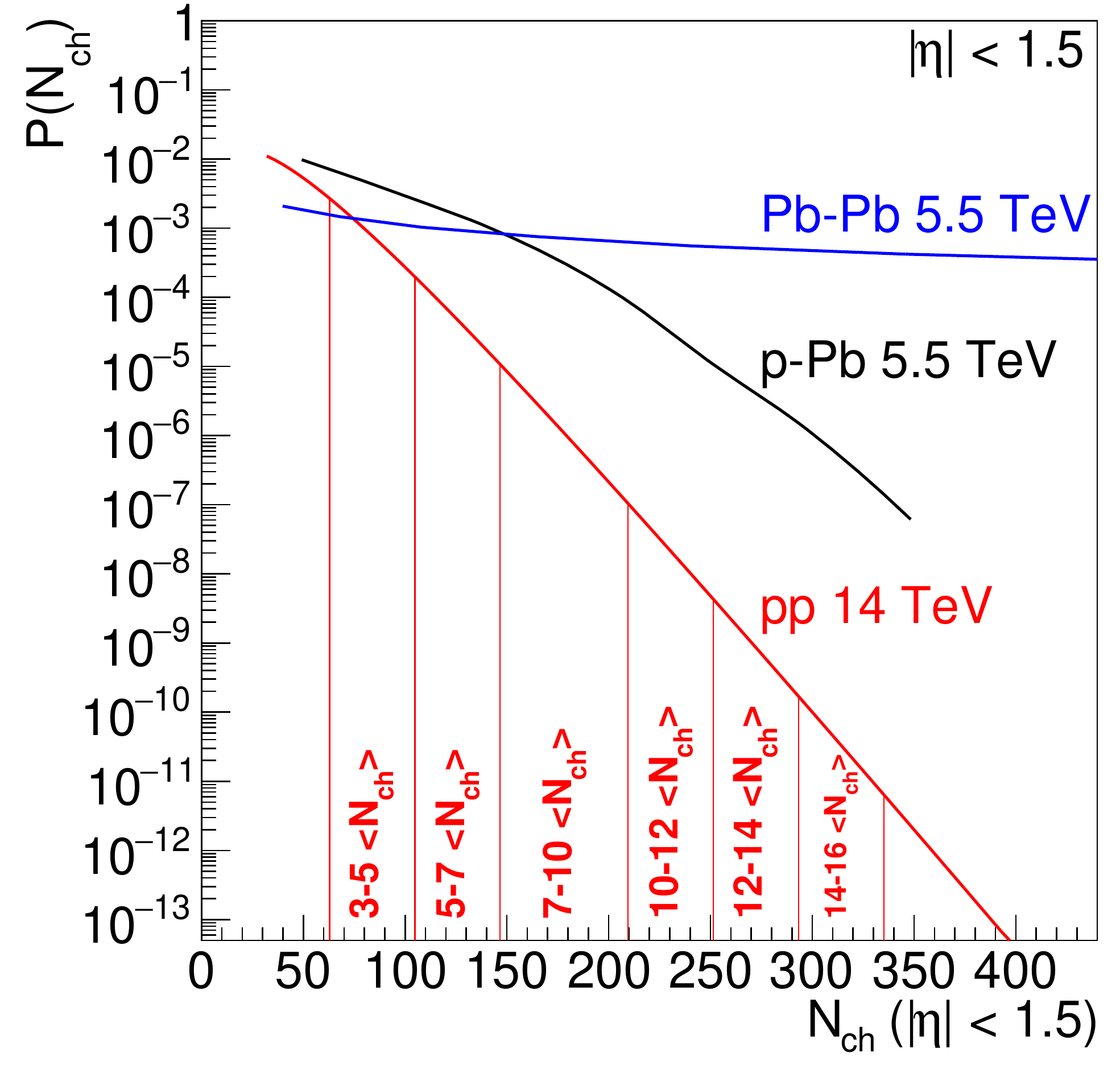}
\caption{Left: Anisotropy as a function of system size in a single-hit scenario compared to the full transport calculation. Figure from Ref.~\cite{Kurkela:2018qeb}. Right: Extrapolated multiplicity distribution in pp collisions at \unit[14]{TeV} compared to p--Pb and Pb--Pb collisions. Figure from Ref.~\cite{Citron:2018lsq}.}
\label{fig:kinetic_future}
\end{figure}

In practice today, hydrodynamics is rather successful in describing the observed phenomena in p--A collisions. While such calculations require in principle local thermal equilibrium, the calculations are quantitatively successful even if the calculations are far from equilibrium (see also Fig.~\ref{fig:aleph_phenix}, right panel) and for large differences between longitudinal and transverse pressure.
Within MC models~\cite{He:2015hfa} and kinetic theory~\cite{Kurkela:2018ygx} it has been shown that few interactions are sufficient to create an anisotropy measurable in the final state. If the system is small enough, the single-hit limit is even close to the full transport~\cite{Kurkela:2018qeb} while there are large deviations for larger systems, see Fig.~\ref{fig:kinetic_future} (left).


The inability of traditional MC codes like PYTHIA to describe the strange baryon production, see Fig.~\ref{fig:strangeness}, and the fact that this cannot be resolved by tuning~\cite{Sjostrand:2018xcd}, have prompted work to extend the baryon production mechanisms considered. Mechanisms that connect the colour flow from different parton--parton interactions (as used in PYTHIA and DIPSY~\cite{Bierlich:2015rha}) or an explicit collective expansion (as used for example in EPOS~\cite{Pierog:2013ria}) bring the models closer to the data, but are also not yet satisfactory~\cite{Acharya:2018orn}. Recently, a promising attempt is to extend PYTHIA into A--A collisions with a model called Angantyr~\cite{Bierlich:2018xfw} whose evolution is worth to be closely followed.

\section{Future}

There are very interesting opportunities ahead to study these open questions further, and understand the underlying QCD processes which give rise to the discussed observations.

At the LHC, the next run (2021--2024) will allow to study extremely rare high-multiplicity pp events, illustrated in Fig.~\ref{fig:kinetic_future} (right) for an integrated luminosity of \unit[200]{pb$^{-1}$}. About 25\,000 events with a multiplicity of 14--16 times the average multiplicity are expected, which is higher than the multiplicity of 65\% central Pb--Pb collisions where significant effects associated with QGP formation are observed. Detailed studies of higher-order flow cumulants, the increase of strangeness production and the search for energy-loss signals will become possible. A detailed discussion of the opportunities with this data sample can be found in Ref.~\cite[Chapter 9]{Citron:2018lsq}.

Furthermore, the study of O--O collisions may give important insight into understanding the puzzling absence of parton energy loss. This symmetric collision system allows for a good selection of the collision geometry and has a similar size than p--Pb collisions, and therefore is still large enough to exhibit parton energy loss~\cite{Citron:2018lsq}. This is currently under discussion at LHC and RHIC.

\section{Summary}

The discovery of QGP phenomena in small collision systems have challenged two paradigms: they have challenged the descriptions explaining phenomena in large heavy-ion collisions. \emph{What is the smallest system where they remain valid?} At the same time, the intriguing effects observed in high-multiplicity pp and p--Pb collisions which are not described by state-of-the-art models, have challenged the standard descriptions used in pp collisions. \emph{Can these remain standard?}
These observations make traditional high-energy physics and traditional heavy-ion physics studies grow closer together which is impressively shown by the tremendous experimental and theoretical progress in the last 8 years.
While it is evident that the underlying QCD is the same theory, the aim of future experimental and theoretical work is to either demonstrate that a unified description from $e^+e^-$ and pp collisions to Pb--Pb collisions is feasible, or to show that different mechanisms are justified. A tremendous opportunity for modelling and understanding of the underlying dynamics lies ahead of us.

\renewcommand{\baselinestretch}{0.95}

\bibliography{jfgo,smallsystems}

\end{document}

%% file: commands.tex
\newcommand{\RAA}          {\ensuremath{R_{\rm AA}}}
\newcommand{\RpA}          {\ensuremath{R_{\rm pA}}}

\newcommand{\pt}           {\ensuremath{p_{\mathrm{T}}}}

\newcommand{\com}[1]       {}

%% file: jfgo.bbl
\providecommand{\href}[2]{#2}\begingroup\raggedright\begin{thebibliography}{10}

\bibitem{Schukraft:2017nbn}
J.~Schukraft, \emph{{QM2017: Status and Key open Questions in
  Ultra-Relativistic Heavy-Ion Physics}},
  \href{https://doi.org/10.1016/j.nuclphysa.2017.05.036}{\emph{Nucl. Phys.}
  {\bfseries A967} (2017) 1}
  [\href{https://arxiv.org/abs/1705.02646}{{\ttfamily 1705.02646}}].

\bibitem{Abelev:2012ola}
{\scshape ALICE} collaboration, \emph{{Long-range angular correlations on the
  near and away side in \pPb\ collisions at $\sNN=5.02$ TeV}},
  \href{https://doi.org/10.1016/j.physletb.2013.01.012}{\emph{Phys. Lett.}
  {\bfseries B719} (2013) 29}
  [\href{https://arxiv.org/abs/1212.2001}{{\ttfamily 1212.2001}}].

\bibitem{Aad:2012gla}
{\scshape ATLAS} collaboration, \emph{{Observation of associated near-side and
  away-side long-range correlations in $\sNN=5.02$ TeV \pPb\ collisions with
  the ATLAS detector}},
  \href{https://doi.org/10.1103/PhysRevLett.110.182302}{\emph{Phys. Rev. Lett.}
  {\bfseries 110} (2013) 182302}
  [\href{https://arxiv.org/abs/1212.5198}{{\ttfamily 1212.5198}}].

\bibitem{Khachatryan:2010gv}
{\scshape CMS} collaboration, \emph{{Observation of Long-Range Near-Side
  Angular Correlations in Proton-Proton Collisions at the LHC}},
  \href{https://doi.org/10.1007/JHEP09(2010)091}{\emph{JHEP} {\bfseries 09}
  (2010) 091} [\href{https://arxiv.org/abs/1009.4122}{{\ttfamily 1009.4122}}].

\bibitem{Citron:2018lsq}
Z.~Citron et~al., \emph{{Future physics opportunities for high-density QCD at
  the LHC with heavy-ion and proton beams}},  in \emph{{HL/HE-LHC Workshop:
  Workshop on the Physics of HL-LHC, and Perspectives at HE-LHC Geneva,
  Switzerland, June 18-20, 2018}}, 2018,
  \href{https://arxiv.org/abs/1812.06772}{{\ttfamily 1812.06772}}.

\bibitem{Loizides:2016tew}
C.~Loizides, \emph{{Experimental overview on small collision systems at the
  LHC}}, \href{https://doi.org/10.1016/j.nuclphysa.2016.04.022}{\emph{Nucl.
  Phys.} {\bfseries A956} (2016) 200}
  [\href{https://arxiv.org/abs/1602.09138}{{\ttfamily 1602.09138}}].

\bibitem{CMS:2012qk}
{\scshape CMS} collaboration, \emph{{Observation of long-range near-side
  angular correlations in pPb collisions at the LHC}},
  \href{https://doi.org/10.1016/j.physletb.2012.11.025}{\emph{Phys. Lett.}
  {\bfseries B718} (2013) 795}
  [\href{https://arxiv.org/abs/1210.5482}{{\ttfamily 1210.5482}}].

\bibitem{Aamodt:2010pa}
{\scshape ALICE} collaboration, \emph{{Elliptic flow of charged particles in
  \PbPb\ collisions at 2.76 TeV}},
  \href{https://doi.org/10.1103/PhysRevLett.105.252302}{\emph{Phys. Rev. Lett.}
  {\bfseries 105} (2010) 252302}
  [\href{https://arxiv.org/abs/1011.3914}{{\ttfamily 1011.3914}}].

\bibitem{Ollitrault:1992bk}
J.-Y. Ollitrault, \emph{{Anisotropy as a signature of transverse collective
  flow}}, \href{https://doi.org/10.1103/PhysRevD.46.229}{\emph{Phys. Rev.}
  {\bfseries D46} (1992) 229}.

\bibitem{ALICE:2017jyt}
{\scshape ALICE} collaboration, \emph{{Enhanced production of multi-strange
  hadrons in high-multiplicity proton-proton collisions}},
  \href{https://doi.org/10.1038/nphys4111}{\emph{Nature Phys.} {\bfseries 13}
  (2017) 535} [\href{https://arxiv.org/abs/1606.07424}{{\ttfamily
  1606.07424}}].

\bibitem{Sjostrand:2007gs}
T.~Sjostrand, S.~Mrenna and P.~Z. Skands, \emph{{A Brief Introduction to PYTHIA
  8.1}}, \href{https://doi.org/10.1016/j.cpc.2008.01.036}{\emph{Comput. Phys.
  Commun.} {\bfseries 178} (2008) 852}
  [\href{https://arxiv.org/abs/0710.3820}{{\ttfamily 0710.3820}}].

\bibitem{Sjostrand:2018xcd}
T.~Sjöstrand, \emph{{Collective Effects: the viewpoint of HEP MC codes}},
  \href{https://doi.org/10.1016/j.nuclphysa.2018.11.010}{\emph{Nucl. Phys.}
  {\bfseries A982} (2019) 43}
  [\href{https://arxiv.org/abs/1808.03117}{{\ttfamily 1808.03117}}].

\bibitem{Alver:2010gr}
B.~Alver and G.~Roland, \emph{{Collision geometry fluctuations and triangular
  flow in heavy-ion collisions}},
  \href{https://doi.org/10.1103/PhysRevC.82.039903,
  10.1103/PhysRevC.81.054905}{\emph{Phys. Rev.} {\bfseries C81} (2010) 054905}
  [\href{https://arxiv.org/abs/1003.0194}{{\ttfamily 1003.0194}}].

\bibitem{ALICE:2011ab}
{\scshape ALICE} collaboration, \emph{{Higher harmonic anisotropic flow
  measurements of charged particles in \PbPb\ collisions at $\sNN=2.76$ TeV}},
  \href{https://doi.org/10.1103/PhysRevLett.107.032301}{\emph{Phys. Rev. Lett.}
  {\bfseries 107} (2011) 032301}
  [\href{https://arxiv.org/abs/1105.3865}{{\ttfamily 1105.3865}}].

\bibitem{Khachatryan:2016txc}
{\scshape CMS} collaboration, \emph{{Evidence for collectivity in pp collisions
  at the LHC}},
  \href{https://doi.org/10.1016/j.physletb.2016.12.009}{\emph{Phys. Lett.}
  {\bfseries B765} (2017) 193}
  [\href{https://arxiv.org/abs/1606.06198}{{\ttfamily 1606.06198}}].

\bibitem{Aad:2019aol}
{\scshape ATLAS} collaboration, \emph{{Measurement of azimuthal anisotropy of
  muons from charm and bottom hadrons in $pp$ collisions at $\sqrt{s}=13$ TeV
  with the ATLAS detector}},
  \href{https://arxiv.org/abs/1909.01650}{{\ttfamily 1909.01650}}.

\bibitem{Acharya:2018dxy}
{\scshape ALICE} collaboration, \emph{{Azimuthal Anisotropy of Heavy-Flavor
  Decay Electrons in $p$-Pb Collisions at $ \sqrt{s_{\rm NN}}$ = 5.02 TeV}},
  \href{https://doi.org/10.1103/PhysRevLett.122.072301}{\emph{Phys. Rev. Lett.}
  {\bfseries 122} (2019) 072301}
  [\href{https://arxiv.org/abs/1805.04367}{{\ttfamily 1805.04367}}].

\bibitem{Sirunyan:2018kiz}
{\scshape CMS} collaboration, \emph{{Observation of prompt J/$\psi$ meson
  elliptic flow in high-multiplicity pPb collisions at $\sqrt{s_\mathrm{NN}} =$
  8.16 TeV}}, \href{https://doi.org/10.1016/j.physletb.2019.02.018}{\emph{Phys.
  Lett.} {\bfseries B791} (2019) 172}
  [\href{https://arxiv.org/abs/1810.01473}{{\ttfamily 1810.01473}}].

\bibitem{Badea:2019vey}
A.~Badea, A.~Baty, P.~Chang, G.~M. Innocenti, M.~Maggi, C.~Mcginn et~al.,
  \emph{{Measurements of two-particle correlations in $e^+e^-$ collisions at 91
  GeV with ALEPH archived data}},
  \href{https://arxiv.org/abs/1906.00489}{{\ttfamily 1906.00489}}.

\bibitem{PHENIX:2018lia}
{\scshape PHENIX} collaboration, \emph{{Creation of quark–gluon plasma
  droplets with three distinct geometries}},
  \href{https://doi.org/10.1038/s41567-018-0360-0}{\emph{Nature Phys.}
  {\bfseries 15} (2019) 214}
  [\href{https://arxiv.org/abs/1805.02973}{{\ttfamily 1805.02973}}].

\bibitem{Habich:2014jna}
M.~Habich, J.~L. Nagle and P.~Romatschke, \emph{{Particle spectra and HBT radii
  for simulated central nuclear collisions of C + C, Al + Al, Cu + Cu, Au + Au,
  and Pb + Pb from $\sqrt{s}=62.4$ - $2760$ GeV}},
  \href{https://doi.org/10.1140/epjc/s10052-014-3206-7}{\emph{Eur. Phys. J.}
  {\bfseries C75} (2015) 15} [\href{https://arxiv.org/abs/1409.0040}{{\ttfamily
  1409.0040}}].

\bibitem{Shen:2016zpp}
C.~Shen, J.-F. Paquet, G.~S. Denicol, S.~Jeon and C.~Gale, \emph{{Collectivity
  and electromagnetic radiation in small systems}},
  \href{https://doi.org/10.1103/PhysRevC.95.014906}{\emph{Phys. Rev.}
  {\bfseries C95} (2017) 014906}
  [\href{https://arxiv.org/abs/1609.02590}{{\ttfamily 1609.02590}}].

\bibitem{Mace:2018vwq}
M.~Mace, V.~V. Skokov, P.~Tribedy and R.~Venugopalan, \emph{{Hierarchy of
  Azimuthal Anisotropy Harmonics in Collisions of Small Systems from the Color
  Glass Condensate}}, \href{https://doi.org/10.1103/PhysRevLett.123.039901,
  10.1103/PhysRevLett.121.052301}{\emph{Phys. Rev. Lett.} {\bfseries 121}
  (2018) 052301} [\href{https://arxiv.org/abs/1805.09342}{{\ttfamily
  1805.09342}}].

\bibitem{Acharya:2018qsh}
{\scshape ALICE} collaboration, \emph{{Transverse momentum spectra and nuclear
  modification factors of charged particles in pp, p-Pb and Pb-Pb collisions at
  the LHC}}, \href{https://doi.org/10.1007/JHEP11(2018)013}{\emph{JHEP}
  {\bfseries 11} (2018) 013}
  [\href{https://arxiv.org/abs/1802.09145}{{\ttfamily 1802.09145}}].

\bibitem{Khachatryan:2015xaa}
{\scshape CMS} collaboration, \emph{{Nuclear Effects on the Transverse Momentum
  Spectra of Charged Particles in pPb Collisions at $\sqrt{s_{_\mathrm {NN}}}
  =5.02$ TeV}},
  \href{https://doi.org/10.1140/epjc/s10052-015-3435-4}{\emph{Eur. Phys. J.}
  {\bfseries C75} (2015) 237}
  [\href{https://arxiv.org/abs/1502.05387}{{\ttfamily 1502.05387}}].

\bibitem{Adam:2016jfp}
{\scshape ALICE} collaboration, \emph{{Centrality dependence of charged jet
  production in p-Pb collisions at $\sqrt{s_\mathrm{NN}}$ = 5.02 TeV}},
  \href{https://doi.org/10.1140/epjc/s10052-016-4107-8}{\emph{Eur. Phys. J.}
  {\bfseries C76} (2016) 271}
  [\href{https://arxiv.org/abs/1603.03402}{{\ttfamily 1603.03402}}].

\bibitem{Sirunyan:2016fcs}
{\scshape CMS} collaboration, \emph{{Measurements of the charm jet cross
  section and nuclear modification factor in pPb collisions at
  $\sqrt{{s}_{NN}}$ = 5.02 TeV}},
  \href{https://doi.org/10.1016/j.physletb.2017.06.053}{\emph{Phys. Lett.}
  {\bfseries B772} (2017) 306}
  [\href{https://arxiv.org/abs/1612.08972}{{\ttfamily 1612.08972}}].

\bibitem{Aaij:2017gcy}
{\scshape LHCb} collaboration, \emph{{Study of prompt D$^{0}$ meson production
  in $p$Pb collisions at $ \sqrt{s_{\mathrm{NN}}}=5 $ TeV}},
  \href{https://doi.org/10.1007/JHEP10(2017)090}{\emph{JHEP} {\bfseries 10}
  (2017) 090} [\href{https://arxiv.org/abs/1707.02750}{{\ttfamily
  1707.02750}}].

\bibitem{Acharya:2019mno}
{\scshape ALICE} collaboration, \emph{{Measurement of prompt D$^0$, D$^+$,
  D$^{*+}$, and D$^+_s$ production in p$-$Pb collisions at
  $\mathbf{\sqrt{{\textit s}_{\rm NN}}~=~5.02~TeV}$}},
  \href{https://arxiv.org/abs/1906.03425}{{\ttfamily 1906.03425}}.

\bibitem{Aaboud:2018quy}
{\scshape ATLAS} collaboration, \emph{{Prompt and non-prompt $J/\psi $ and
  $\psi (2\mathrm {S})$ suppression at high transverse momentum in
  $5.02~\mathrm {TeV}$ Pb+Pb collisions with the ATLAS experiment}},
  \href{https://doi.org/10.1140/epjc/s10052-018-6219-9}{\emph{Eur. Phys. J.}
  {\bfseries C78} (2018) 762}
  [\href{https://arxiv.org/abs/1805.04077}{{\ttfamily 1805.04077}}].

\bibitem{Aaij:2019lkm}
{\scshape LHCb} collaboration, \emph{{Measurement of $B^+$, $B^0$ and
  $\Lambda_b^0$ production in $p\mkern 1mu\mathrm{Pb}$ collisions at
  $\sqrt{s_\mathrm{NN}}=8.16\,{\rm TeV}$}},
  \href{https://doi.org/10.1103/PhysRevD.99.052011}{\emph{Phys. Rev.}
  {\bfseries D99} (2019) 052011}
  [\href{https://arxiv.org/abs/1902.05599}{{\ttfamily 1902.05599}}].

\bibitem{Acharya:2017okq}
{\scshape ALICE} collaboration, \emph{{Constraints on jet quenching in p-Pb
  collisions at $\mathbf{\sqrt{s_{NN}}}$ = 5.02 TeV measured by the
  event-activity dependence of semi-inclusive hadron-jet distributions}},
  \href{https://doi.org/10.1016/j.physletb.2018.05.059}{\emph{Phys. Lett.}
  {\bfseries B783} (2018) 95}
  [\href{https://arxiv.org/abs/1712.05603}{{\ttfamily 1712.05603}}].

\bibitem{Acharya:2018njl}
{\scshape ALICE} collaboration, \emph{{Analysis of the apparent nuclear
  modification in peripheral Pb–Pb collisions at 5.02 TeV}},
  \href{https://doi.org/10.1016/j.physletb.2019.04.047}{\emph{Phys. Lett.}
  {\bfseries B793} (2019) 420}
  [\href{https://arxiv.org/abs/1805.05212}{{\ttfamily 1805.05212}}].

\bibitem{Morsch:2017brb}
C.~Loizides and A.~Morsch, \emph{{Absence of jet quenching in peripheral
  nucleus–nucleus collisions}},
  \href{https://doi.org/10.1016/j.physletb.2017.09.002}{\emph{Phys. Lett.}
  {\bfseries B773} (2017) 408}
  [\href{https://arxiv.org/abs/1705.08856}{{\ttfamily 1705.08856}}].

\bibitem{Aaboud:2017acw}
{\scshape ATLAS} collaboration, \emph{{Measurement of multi-particle azimuthal
  correlations in $pp$, $p+$Pb and low-multiplicity Pb$+$Pb collisions with the
  ATLAS detector}},
  \href{https://doi.org/10.1140/epjc/s10052-017-4988-1}{\emph{Eur. Phys. J.}
  {\bfseries C77} (2017) 428}
  [\href{https://arxiv.org/abs/1705.04176}{{\ttfamily 1705.04176}}].

\bibitem{He:2015hfa}
L.~He, T.~Edmonds, Z.-W. Lin, F.~Liu, D.~Molnar and F.~Wang, \emph{{Anisotropic
  parton escape is the dominant source of azimuthal anisotropy in transport
  models}}, \href{https://doi.org/10.1016/j.physletb.2015.12.051}{\emph{Phys.
  Lett.} {\bfseries B753} (2016) 506}
  [\href{https://arxiv.org/abs/1502.05572}{{\ttfamily 1502.05572}}].

\bibitem{Kurkela:2018ygx}
A.~Kurkela, U.~A. Wiedemann and B.~Wu, \emph{{Nearly isentropic flow at
  sizeable $\eta/s$}},
  \href{https://doi.org/10.1016/j.physletb.2018.06.064}{\emph{Phys. Lett.}
  {\bfseries B783} (2018) 274}
  [\href{https://arxiv.org/abs/1803.02072}{{\ttfamily 1803.02072}}].

\bibitem{Kurkela:2018qeb}
A.~Kurkela, U.~A. Wiedemann and B.~Wu, \emph{{Opacity dependence of elliptic
  flow in kinetic theory}},
  \href{https://doi.org/10.1140/epjc/s10052-019-7262-x}{\emph{Eur. Phys. J.}
  {\bfseries C79} (2019) 759}
  [\href{https://arxiv.org/abs/1805.04081}{{\ttfamily 1805.04081}}].

\bibitem{Bierlich:2015rha}
C.~Bierlich and J.~R. Christiansen, \emph{{Effects of color reconnection on
  hadron flavor observables}},
  \href{https://doi.org/10.1103/PhysRevD.92.094010}{\emph{Phys. Rev.}
  {\bfseries D92} (2015) 094010}
  [\href{https://arxiv.org/abs/1507.02091}{{\ttfamily 1507.02091}}].

\bibitem{Pierog:2013ria}
T.~Pierog, I.~Karpenko, J.~M. Katzy, E.~Yatsenko and K.~Werner, \emph{{EPOS
  LHC: Test of collective hadronization with data measured at the CERN Large
  Hadron Collider}},
  \href{https://doi.org/10.1103/PhysRevC.92.034906}{\emph{Phys. Rev.}
  {\bfseries C92} (2015) 034906}
  [\href{https://arxiv.org/abs/1306.0121}{{\ttfamily 1306.0121}}].

\bibitem{Acharya:2018orn}
{\scshape ALICE} collaboration, \emph{{Multiplicity dependence of light-flavor
  hadron production in pp collisions at $\sqrt{s}$ = 7 TeV}},
  \href{https://doi.org/10.1103/PhysRevC.99.024906}{\emph{Phys. Rev.}
  {\bfseries C99} (2019) 024906}
  [\href{https://arxiv.org/abs/1807.11321}{{\ttfamily 1807.11321}}].

\bibitem{Bierlich:2018xfw}
C.~Bierlich, G.~Gustafson, L.~Lönnblad and H.~Shah, \emph{{The Angantyr model
  for Heavy-Ion Collisions in PYTHIA8}},
  \href{https://doi.org/10.1007/JHEP10(2018)134}{\emph{JHEP} {\bfseries 10}
  (2018) 134} [\href{https://arxiv.org/abs/1806.10820}{{\ttfamily
  1806.10820}}].

\end{thebibliography}\endgroup
